\documentstyle[twocolumn,prl,aps]{revtex}

\tighten
\begin{document}

\draft

\title{Star-shaped Local Density of States around Vortices
             in a Type II Superconductor}

\author{Nobuhiko Hayashi,\cite{haya} 
Masanori Ichioka,\cite{oka} and Kazushige Machida\cite{machida}}
\address{Department of Physics, Okayama University,\\
         Okayama 700, Japan\medskip\\
\parbox{14cm}{\rm
     The electronic structure of vortices in a type II superconductor is 
analyzed within the quasi-classical Eilenberger framework.
     The possible origin of a sixfold ``star'' shape of the local density
of states, observed by scanning tunneling microscope experiments on
NbSe$_2$, is examined
in the light of the three effects; the anisotropic pairing,
the vortex lattice,
and the anisotropic density of states at the Fermi surface.
     Outstanding features of split parallel rays of this
star are well explained in terms of an anisotropic $s$-wave pairing.
This reveals a rich internal electronic structure associated with a vortex
core.}}

%\end{abstract}
\maketitle

%%%%%%%%%%%%%%%%%%%%%%%%%%%%%%%%%%%%%%%%
     Much attention has been focused on a vortex structure in high $T_c$ 
cuprate superconductors, yet some of fundamental
problems associated with vortices still remain even in 
conventional superconductors.
     The electronic structure around a vortex core
is one of them.
     Notably,
Hess {\it et al.}\cite{hess1,hess1-1,hess2,hess3,hess4,book} have
done a series of remarkable and beautiful scanning tunneling microscope
(STM) experiments on a layered
hexagonal compound 2H-NbSe$_2$ ($T_c$=7.3K) to reveal the detailed
spatially resolved electronic structure around a vortex core.
     They provide direct and vivid images of individual vortices
and the flux line lattice.
     It was shown theoretically that the STM images
observed by Hess {\it et al.} and Renner {\it et al.}\cite{fischer}
reflect the local density of states (LDOS) for bound states
around a vortex.
On the basis of the Bogoliubov-de Gennes (BdG) theory\cite{shore,gygi0,gygi}
or the quasi-classical Eilenberger (QCE) theory\cite{klein1-1,ullah},
the case of the cylindrical symmetric isolated vortex was considered there.
     However, the detailed structure of the LDOS observed is not fully
understood.

     The outstanding features
of the experimental findings\cite{hess1-1,hess2,hess3,hess4,book}
are summarized as follows when the magnetic field $H$ is applied 
perpendicular to the hexagonal plane:
     (1) The LDOS for quasiparticle excitations has a sixfold
``star'' shape centered at a core.
     (2) The orientation of this star depends on the bias energy.
     At zero bias the ``ray'' extends away from the nearest-neighbor vortex
direction [Fig. \ref{fig:1}(a)] 
where the conventional 60$^{\circ}$ Abrikosov vortex lattice is formed. 
     Upon increasing the bias voltage the star rotates
by 30$^{\circ}$ [Fig. \ref{fig:1}(c)]. 
     (3) In the intermediate bias voltage a ray splits into
a pair of nearly parallel rays,
keeping its direction fixed [Fig. \ref{fig:1}(b)].
     (4) In the spectral evolution as shown in Fig. 9 in Ref.\cite{book}
(also see Fig. 6 in Ref.\cite{hess2})
there exist additional inner ridges other than 
the outer ridges evolving from the zero bias peak at a core into the
bulk BCS like gap edges far from a core.

     To understand these experimental
results,
Gygi and Schl\"uter\cite{gygi,gygi2}
discussed the sixfold symmetric structure on the basis of 
the BdG theory by introducing a sixfold symmetric perturbation term 
and using their numerical solutions of the cylindrical symmetric case.
     While they explained the above mentioned features (1) and (2), 
the origin of their perturbation term is not clear, and it is
uncertain whether the sixfold symmetric term can be treated 
by a perturbation theory.
     We notice that the features (3) and (4) are
difficult to be explained in their framework.
     In the following, we show that an anisotropic pairing (or gap) effect
which has been ignored so far
is crucial for the bound states around a vortex.

     The purpose of this paper is to clarify 
the origin of the characteristic star-shaped LDOS
by the QCE theory\cite{eilen}. 
     The QCE approach is more suitable than the BdG approach
for the case cylindrical symmetry is broken.
     The anisotropic pairing can be treated easily by the QCE theory.
     As for the origin of the sixfold symmetric structure, 
the following possibilities are enumerated; 
(A) the effect of an anisotropic pairing, 
(B) the effect of nearest-neighbor vortices,
that is, the effect of the vortex lattice, and
(C) the effect of the anisotropic density of states
at the Fermi surface (FS).
     We calculate the LDOS for each case (A)-(C), and investigate
which case can explain the LDOS of the STM experiments.

     We introduce an anisotropic $s$-wave pairing 
$\Delta({\bf r},{\bf k})=
\Delta_0({\bf r}) F(\theta)=
\Delta_0({\bf r})(1+ {\it c}_{\rm A}\cos6\theta)$ with hexagonal
symmetry and assume two-dimensional circular FS,
neglecting a small warping along the hexagonal
$z$ axis ($\parallel {\bf H}$), which is appropriate to NbSe$_2$.
     Here ${\bf r}=(x, y)$ is the center of mass coordinate and
${\bf k}$ the relative coordinate of a Cooper pair.
     Now ${\bf k}$ is denoted
by an angle $\theta$ measured from the $a$ axis (or $x$ axis)
in the hexagonal plane. 
     Thus the parameter ${\it c}_{\rm A}$ denotes the degree of anisotropy
in a pairing\cite{marko,clem}.
     The case ${\it c}_{\rm A}=0$ corresponds to
a conventional isotropic pairing.

%%%%%
     Our calculation is performed after the method of Ref. \cite{ichioka}
for the isolated vortex and that of Klein\cite{klein2,klein}
for the vortex lattice.
     We consider a transport-like 
Eilenberger equation\cite{eilen} for the
quasi-classical Green 
function $\hat{g}({\bf r}, 
\theta,i\omega_n)=-i\pi\pmatrix{g & if\cr -if^{\dagger} &-g \cr}$
in a 2$\times$2 matrix form, namely,
%%%
\begin{eqnarray}
i&{\bf v}_{\rm F}&\cdot\nabla\hat{g}({\bf r}, \theta,i\omega_n)\nonumber\\
& & { }+ [\pmatrix{i\omega_n & -\Delta({\bf r},\theta)\cr
\Delta^*({\bf r},\theta)
&-i\omega_n\cr}, \  \ \hat{g}({\bf r}, \theta,i\omega_n)]=0,
\end{eqnarray}
supplemented by the normalization condition:
$\hat{g}({\bf r},\theta,i\omega_n)^2
=-\pi^2\hat{1}$.
     The Fermi velocity is $v_{\rm F}$.
     The bracket $[\ \ , \ \ ]$ is a commutator.
     The self-consistent equation is given by 
%%%
\begin{equation}
\Delta({\bf r},\theta)=2\pi T\sum_{\omega_n>0}\int{d\theta'\over 2\pi}
\rho (\theta')V(\theta,\theta')f({\bf r}, \theta',i\omega_n),
\end{equation}
where the pairing interaction
$V(\theta,\theta')=vF(\theta)F(\theta')$ is assumed to be separable.
     The $\theta$-dependent density of states at the FS
is $\rho (\theta)$.
     We calculate $\Delta_0({\bf r})$ self-consistently from Eqs. (1)
and (2) for $T/T_c=0.1$.
     The LDOS is given by
%%%
\begin{eqnarray}
N({\bf r}, E)&=&\int_0^{2\pi}{d\theta \over 2\pi} \rho (\theta)
N({\bf r}, \theta, E)\nonumber\\
&=&\int_0^{2\pi}{d\theta \over 2\pi} \rho (\theta)
{\rm Re}\ g({\bf r}, \theta, i\omega_n \rightarrow E+i\eta),
\end{eqnarray}
where $\eta$ ($>$0) is an infinitesimal constant. 
     To obtain $ g({\bf r}, \theta, i\omega_n \rightarrow E+i\eta)$, 
we solve Eq. (1) for $\eta-iE$
instead of the Matsubara frequency $\omega_n$ 
using the self-consistently obtained pair potential.
%%%%%
%%%%%
     From now on, the energy $E$ and the length $r$ are scaled by
the uniform gap  $\Delta_0$ at zero temperature
and $\xi=v_{\rm F}/\Delta_0$ respectively.

     Let us start out by considering (A)
the effect of an anisotropic pairing. 
     The anisotropic pairing is suggested
from the STM experiment at zero field. 
     The observed $I$-$V$ tunneling spectrum
in the bulk indicates a substantial gap anisotropy
(the gap amplitude with the averaged value 1.1mV
distributes from 0.7mV to 1.4mV, see Fig.1 in Ref.\cite{hess2}),
which is consistent with the density of states
in the anisotropic pairing case (see Fig. 5 in Ref. \cite{clem}).
     To see the influence of (A) clearly,
we neglect the other effects (B) and (C) here.
     We calculate the LDOS
in the isolated vortex case assuming an isotropic FS.

     In Figs. 1(d), 1(e), and 1(f), we show the calculated LDOS for several
bias energies $E$ in the case ${\it c}_{\rm A}=0.5$.
     It is seen from Fig. \ref{fig:1}(d) that the sixfold star
centered at a core is oriented away from the $x$ axis by 30$^{\circ}$
for $E$=0.
     The rays along the $30^\circ$ direction can be seen to extend up to
$r\sim 3$ from the center.
      It corresponds to the following fact. 
      As for the bound state around a vortex core at $E=0$,
the wave function of 
the quasiparticle running through the vortex center has the maximum 
amplitude. 
      Among them, because $F(\theta)$ is most depressed 
at $\theta = 30^\circ$, the quasiparticle with the  momentum ${\bf k}$
along the 
$30^\circ$ direction feels weakest
superconducting pair potential, 
and its wave function keeps large amplitude until far from the vortex.

     It is seen from Fig. \ref{fig:1}(e) that at the intermediate energy
each ray splits into two parallel rays,
keeping its direction.
      This characteristic feature
is precisely observed by Hess [Fig. \ref{fig:1}(b)] (also see Refs.
\cite{hess3,hess4})
and is absent in the result by Gygi and Schl\"uter\cite{gygi,gygi2}.
     For this intermediate excitation energy, the wave function of the
quasiparticle passing at a small finite distance from the vortex center
has the maximum amplitude.
     Among them, the quasiparticle with the momentum ${\bf k}$
along the $30^\circ$
and its equivalent directions keeps large amplitude
until far from the vortex, and forms the nearly parallel structure in 
the ${\bf r}$ space.

     With increasing the bias energy $E$ further, the sixfold star
becomes a more extended one,
and its orientation rotates by 30$^{\circ}$
as seen from Fig. \ref{fig:1}(f).
      Note that the heads of each ray spread out.
      It coincides with 
the observation in Fig. \ref{fig:1}(c).

     Another way to examine the quasiparticle excitations 
in the vortex state is to see how the spectrum evolves
along the radial lines.
     The calculated spectrum evolution shown in Fig. \ref{fig:2} is
compared with the data in Fig. 9 in Ref.\cite{book}
(also see Fig. 6 in Ref.\cite{hess2}). 
     Here the radial line is chosen to make an angle 15$^{\circ}$
from the $x$ axis, where there are five ridges for $E>0$. 
     We label these ridges as $\alpha$-$\varepsilon$. 
     As $r$ increases, two higher energy ridges $\alpha$ and $\beta$
approach the maximum 
of the gap energy $1+{\it c}_{\rm A}$,
and the ridge $\gamma$ is found to become very 
weak intensity. 
     The ridge $\delta$, which approaches 
the minimum of the gap energy $1-{\it c}_{\rm A}$ with increasing $r$,
corresponds to that reported in the isotropic pairing 
model by Gygi and Schl\"uter (see Fig. 15 in Ref.\cite{gygi}). 
     The ridge $\varepsilon$ is the result reflecting
the sixfold symmetric ray structure.
     The ridges $\delta$ and $\varepsilon$ correspond to
the outer and inner ridges in the experimental data\cite{hess2,book}
by Hess {\it et al.}, respectively.

     To clarify the behavior of these ridges,
we show in Fig. \ref{fig:3} the trace of 
the ridges in the LDOS as a function of $r$ along the radial lines for 
$0^\circ$ (a), $15^\circ$ (b), and $30^\circ$ (c) from the $x$ axis. 
     For comparison, the experimental data\cite{hess2} are also presented. 
     We direct our attention to the inner ridge (line $\varepsilon$), 
which strongly depends on the radial angle.
     It moves toward the lower energy side as the angle increases. 
     At $30^\circ$, it reduces to the ridge
at $E=0$ (horizontal $r$-axis). 
     This behavior of the inner ridge agrees well with that of the 
experimental data.

     Thus, the effect of the anisotropic pairing
can explain not only (1) the sixfold 
star shape and (2) the $30^\circ$ rotation, but also 
(3) the split parallel ray structure at the intermediate energy and 
(4) the characteristic behavior of the inner ridges in the spectral 
evolution.

     Notice that this characteristic spreading of the ray structure
depends on the degree of anisotropy ${\it c}_{\rm A}$;
     In the isotropic limit (${\it c}_{\rm A}=0$) the star shape turns into
a circular object.
     As ${\it c}_{\rm A}$ increases,
the star structure spreads out farther from a core.
     This implies that in NbSe$_2$ a substantial gap anisotropy
must exist even within the hexagonal plane.
     It is consistent with the tunneling experiment\cite{hess2}
in the bulk mentioned earlier.

     Let us argue a possible origin of this rather strong gap anisotropy.
     It is known that the system enters into a charge density wave (CDW)
state at 32K which coexists with superconductivity at lower temperatures.
     The FS's consist of two kinds of cylinders,
each centered at the $\Gamma$ point and $K$ point in reciprocal space
\cite{harima}.
     The CDW reorganizes the FS's,
opening up the CDW gaps by nesting.
     The remaining FS's left over by the CDW gap formation
are available for the superconductivity.
     Therefore, we can expect a substantial gap anisotropy
within the hexagonal plane.

%%%%%
     Next, we turn to (B) the effect of the vortex lattice. 
     The LDOS is calculated in the case of a triangular vortex lattice 
by using the self-consistently obtained pair potential. 
 Here, neglecting the other effects (A) and (C), we consider the 
isotropic $s$-wave pairing (${\it c}_{\rm A}=0$) and assume an isotropic 
FS.
     As for the LDOS in the vortex lattice case,
Klein calculated only the momentum-resolved LDOS $N({\bf r}, \theta, E)$
in Eq. (3)
for specific ${\bf k}$ direction parallel to $m{\bf r}_1+n{\bf r}_2$
($m$ and $n$ are integer,
${\bf r}_1$ and ${\bf r}_2$ are unit vectors of the vortex lattice)
because the so-called symmetric method was used\cite{klein}.
     Using the explosion method\cite{ichioka,klein2,klein},
we succeed to calculate $N({\bf r}, \theta, E)$
for arbitrary ${\bf k}$ direction.
     Then we can obtain the LDOS integrated over ${\bf k}$
directions.

     Important results obtained for the vortex lattice are;
     (i) We do find a sixfold star shape of the LDOS
and the 30$^\circ$ rotation
upon elevating the bias energy.
     The sixfold star originates
from the triangular vortex lattice effect.
     (ii) The orientation of the star coincides
with the STM data, namely the ray extends toward
the next nearest-neighbor vortex direction at lower bias energy.
     Therefore we succeed in
determining the absolute direction relative to the flux lattice.
     This is one of the most eminent features in the STM data and absent in
Gygi and Schl\"uter\cite{gygi,gygi2}.
     (iii) However, the split parallel ray structure at the intermediate 
energy is not reproduced.
     (iv) 
 The characteristic sixfold symmetric LDOS appears only at a high 
magnetic field such as 1 Tesla for the material parameters appropriate  
to ${\rm NbSe_2}$ (The BCS coherence length 77\AA\ and the BCS 
penetration depth 690\AA\cite{hess2}),
where the core regions substantially 
overlap each other. 
 At a lower magnetic field such as 0.1 Tesla, the LDOS reduces to the 
almost circular structure.

     As for (C) the  effect of the anisotropic density of states
at the FS,
we find that reflecting the assumed sixfold anisotropy
the LDOS exhibits the characteristic star shape and
the 30$^{\circ}$ rotation even for isotropic $s$-pairing.
However, the split parallel ray structure at the intermediate energy
is not reproduced.

     In conclusion,
by solving the quasi-classical Eilenberger equation self-consistently,
we have examined the possible origins of the star-shaped LDOS
observed in the STM experiments on NbSe$_2$.
    Among the three effects (A), (B), and (C), each one can 
reproduce the experimental features (1) the sixfold 
star shape and (2) the $30^\circ$ rotation. 
    In ${\rm NbSe_2}$, all the three effects may play important roles. 
    However, to reproduce the features (3) the split parallel ray 
structure at the intermediate energy and (4) the characteristic 
behavior of the inner ridges in the spectral evolution, we have to 
consider the LDOS by including the effect of the anisotropic pairing.

    As for the relation between the orientation of the star shape 
and that of the vortex lattice, 
the calculation for the vortex lattice
has to be done by including the anisotropy effect. 
     It is noted that the orientations of the star shape
and of the vortex lattice relative to the underlying crystal lattice
are internally correlated through the gap function.
     The gap function itself is determined by the FS topology.
     The complete answer to this can be obtained
by considering all the three effects simultaneously,
which belongs to a future work.
     The present study underlies the basic understanding for
fundamental electronic properties of vortices
both for conventional and unconventional superconductors,
revealing a rich internal electronic structure associated with vortices.

     In connection to the anisotropic pairing effect,
we point out that high $T_c$ superconductors  should
exhibit similar STM characteristics if a $d$-wave pairing
such as $\Delta({\bf r},{\bf k})=\Delta_0({\bf r}) \cos2\theta$
is realized.
     We predict a ``fourfold'' star and its 45$^{\circ}$ rotation
there\cite{ichioka}.

We would like to thank H. F. Hess for providing unpublished data
and the original photographs for reproducion, and 
for encouragement, which certainly motivates the present 
project.

%%%%%%%%%%%%%%%%%%%%%%%%%%%%%%%%%%%%%%%%

%%%%%%%%%%%%%%%%%%%%%%%%%%%%%%%%%%%%%%%%
\begin{figure}
\caption{
     Tunneling conductance images observed
by Hess {\it et al.} at 0.1 Tesla
for the bias voltage 0.0mV (a), 0.24mV (b), 0.48mV (c),
 where 1759\AA\ $\times$ 1759\AA\ is shown (also see Refs. [4,5]). 
     The nearest-neighbor direction of the vortex lattice is
the horizontal direction.
     The LDOS images calculated for $E=0$ (d), 0.2 (e), and 0.32 (f),
where $6\xi \times 6\xi$ is shown.
}
\label{fig:1}
\end{figure}

\begin{figure}
\caption{
     Evolution of the spectra along a radial line
at the angle 15$^{\circ}$ from the $x$ axis.
     Each ridge is labeled as $\alpha$-$\varepsilon$
from the high energy side.
}
\label{fig:2}
\end{figure}

\begin{figure}
\caption{
     The ridge energies of the LDOS as a function of $r$ along 
the radial lines for 0$^{\circ}$ (a), 15$^{\circ}$ (b), and 30$^{\circ}$ (c)
from the $x$ axis (solid lines).
     The case (b) corresponds to Fig. 2, where
the labels $\alpha$-$\varepsilon$ are the same as those in Fig. 2.
     Experimental data[3] are also presented by points $\bullet$ (outer)
and $\circ$ (inner), where $r$ and $E$ are
scaled by 350\AA\ and 1.67mV, respectively.
}
\label{fig:3}
\end{figure}

\end{document}